\documentstyle[12pt]{article}

\newcommand{\be}{\begin{equation}}
\newcommand{\ee}{\end{equation}}
\newcommand{\bea}{\begin{eqnarray}}
\newcommand{\eea}{\end{eqnarray}}
\newcommand{\bean}{\begin{eqnarray*}}
\newcommand{\eean}{\end{eqnarray*}}
\newcommand{\eqn}[1]{(\ref{#1})}

\newcommand{\del}{\partial}
\newcommand{\lp}{Lie--Poisson\ }

\newcommand{\I}{\mbox{\rm I} \hspace{-0.5em} \mbox{\rm I}\,}

\newcommand{\R}{\mbox{I \hspace{-0.82em} R}}

\newcommand{\g}{{\cal G}}

\def\thebibliography#1{\section*{REFERENCES\markboth
 {REFERENCES}{REFERENCES}}\list
 {[\arabic{enumi}]}{\settowidth\labelwidth{[#1]}\leftmargin\labelwidth
 \advance\leftmargin\labelsep
 \usecounter{enumi}}
 \def\newblock{\hskip .11em plus .33em minus -.07em}
 \sloppy
 \sfcode`\.=1000\relax}

\begin{document}

\hfill DSF-T-93/21\\
\hfill hep-th/9308026
\begin{center}
\vskip 2.5cm
{\LARGE \bf Dynamical Aspects of Lie--Poisson Structures}
\vskip 1.0cm
{\Large F.~LIZZI, G.~MARMO, G.~SPARANO and P.~VITALE\footnote{\scriptsize
LIZZI@NA.INFN.IT, GIMARMO@NA.INFN.IT, SPARANO@NA.INFN.IT,
VITALE@NA.INFN.IT}}
\\
{\large  Dipartimento di Scienze Fisiche and I.N.F.N., Sez.\ di Napoli\\
Mostra d'Oltremare Pad.~19, 80125 Napoli.}
\vskip 0.4cm
\end{center}

\begin{abstract}
Quantum Groups can be constructed by applying the quantization by deformation
procedure to Lie groups endowed with a suitable Poisson bracket. Here we try to
develop an understanding of these structures by investigating dynamical systems
which are associated with this bracket. We look at $SU(2)$ and $SU(1,1)$, as
submanifolds of a 4--dimensional phase space with constraints, and deal with
two classes of problems. In the first set of examples we consider some
hamiltonian systems associated with Lie-Poisson structures and we investigate
the equations of the motion. In the second set of examples we consider systems
which preserve the chosen bracket, but are dissipative. However in this
approach, they survive the quantization procedure.
\end{abstract}

\vfill
\begin{center}
May 1993
\end{center}
\newpage
\setcounter{footnote}{1}

\

\section{Introduction}

The presence of a Poisson bracket on a manifold is an important ingredient
to start the quantization procedure of dynamical systems. When the manifold is
 a Lie Group, of
particular interest are Lie--Poisson brackets. In this case these Lie groups
are called \lp groups.
The aim of this paper is to consider \lp groups as carrier spaces of
dynamical systems.
The interest for \lp groups stems from the observation that they may be
regarded as "dequantization" of Quantum Groups; we hope thereof to gain an
understanding of these structures by putting them into a dynamical
perspective.

As it is well known
quantum groups can be seen as non commutative generalizations of topological
spaces which have a group structure; such a structure induces
an abelian Hopf algebra structure on the algebra of smooth functions ${\cal
F}(G)$ defined on the group. Quantum groups are
defined then as non abelian Hopf algebras (\cite{tak,doeb}). A way to
generate
them consists in deforming the product of the abelian Hopf algebra of
functions ${\cal F} (G)$ into a
nonabelian one ($*$-product), using the so called quantization by deformation
procedure or $*$-quantization.
(\cite{tak,doeb,lich}).
This quantization technique gives a deformed product once it is assigned a
Poisson bracket on the algebra ${\cal F}(G)$. In order to obtain that the
deformed algebra is a Hopf algebra, namely a quantum group, the starting
group $G$ has to be endowed with a \lp structure, that is, the group
multiplication, $m ~:~G \times G \rightarrow G$, must be a Poisson map.
 Lie groups which enjoy this property are said \lp groups.
Being our interest mainly concerned with exemplification,
we refer to \cite{tak,doeb,drinf} for the general theory.

We will consider two examples of three dimensional Lie groups, namely
$SU(2)$ and $SU(1,1)$, as carrier spaces of dynamical systems, and we will
look at the group manifold as a 3-dimensional surface in $\R^4$.
We will then consider a Lie-Poisson structure on $G$, that is a bivector
field $\Lambda$ which defines a Poisson bracket on the ${\cal F}(G)$
algebra of smooth functions on $G$
\be
\Lambda (df\wedge dg)=\{ f, g \}~~~~~~~~~~~f, g \in {\cal F}(G);
\ee
in addition, such a bracket is also compatible with
the group structure, being the group multiplication a Poisson map.
Since the group manifold is three dimensional, the Poisson bracket
must be degenerate, that is there must be, at least locally, a Casimir function
${\cal C}_1\in{\cal F}(G)$ such that:
\be
\{ {\cal C}_1, f \} = 0 \quad \forall f \in{\cal F}(G) \ .
\ee
Moreover, when we regard the group $G$ as a three dimensional surface in
$\R^4$, vector fields on $\R^4$ entering in the definition of $\Lambda$,
must be tangent to the group submanifold. If this submanifold is defined by
the level set of a function ${\cal C}_2$, this function will be another
Casimir for our bracket.

\lp structures which we will consider on $SU(2)$ and $SU(1,1)$ are of
the form
\be
\Lambda = r^{ij} (X_i \wedge X_j - \tilde X_i \wedge \tilde X_j)
\label{cob}
\ee
with $r^{ij}\in \R$, $X_i$ the left invariant vector fields, and $\tilde X_j$
the
right invariant ones. In fact for simple Lie groups such a structure
not only satisfies the request of being \lp, but all \lp structures are of
this form (\cite{drinf}).

\section{Hamiltonian Systems}

We now consider the \lp structure defined in equation \eqn{cob}
in some specific cases, and investigate
the dynamics one obtains in the presence of a Hamiltonian. Obviously there is
great arbitrariness in the choice of the Hamiltonian, and in our examples we
have chosen some `natural' ones.

\subsection{$SU(2)$}
We indicate with $y^\mu$, $\mu=0,\ldots,3$ some coordinates in $\R^4$ and
represent $SU(2)$ as unitary $2 \times 2$ complex matrices of the form
\be
g= y^0 \I + i \sigma_j y^j
\ee
where the $\sigma$'s are the Pauli matrices and $\I$ is the identity
matrix. The group manifold is defined by the constraint:
\be
\det g = y_0^2 + y_1^2 + y_2^2 + y_3^2=1 \ \ .
\ee
In order to write the Poisson bracket \eqn{cob} in $\R^4$ we use the following
 realization for left and right invariant vector fields:
\bea
X_i & = & y^0 \del_i - y^i \del_0 + \varepsilon_{ijk} y^j \del_k \nonumber\\
\tilde X_i & = & y^0 \del_i - y^i \del_0 - \varepsilon_{ijk} y^j \del_k\ \ .
\label{V}
\eea

Since the manifold is a 3--sphere all directions are equivalent and therefore
we may take, without loss of generality, the \lp bivector field to be
\be
\Lambda =
{1\over 2} (X_1 \wedge X_2 - \tilde X_1 \wedge \tilde X_2) \label{Lambda} \ \ .
\ee

Using the $\R^4$ representation of the vector fields given by \eqn{V} we may
write the bivector field \eqn{Lambda} in the following way:
\be
\Lambda = (y_3 \del_0 - y_0 \del_3) \wedge y_\mu \del_\mu + \det g \del_0
\wedge \del _3 \label{Lambdacoord}
\ee
with
\be
\det g ={\cal C}_2 = y_\mu y^\mu \label{C2} \ \ .
\ee
This function is in fact one of the two Casimirs of the Poisson bracket
given by $\Lambda$. The other Casimir can easily be recognised to be any
function of the ratio ${\cal C}_1=y_2/y_1$; for future convenience we will
also consider the Casimir one--form related to ${\cal C}_1$, $\alpha _1$,
defined as
\be
\alpha _1= y_2 dy_1 - y_1 d y_2=-y_1^2 d{\cal C}_1\ .\label{C1}
\ee
A 1-form $\alpha$ is said to be a Casimir 1-form for $\Lambda$ if it admits
an integrating factor and $\Lambda (\alpha)=0$. Under some regularity
assumptions, Casimir 1-forms define codimension one submanifolds in the
obvious way. The advantage of Casimir 1-forms with respect to Casimir
functions is relevant when dealing with global properties. For instance
$\alpha_1$ in \eqn{C1} is smooth and well defined over all $\R^2$ while for
${\cal C}_1=y_2/y_1$ we have to remove $y_1=0$ thus getting a disconnected
manifold $(\R -\{0\})\times \R$.
The Casimirs ${\cal C}_2$ and $\alpha_1$ define a two--dimensional surface, say
$\Sigma$, where $\Lambda$ is non degenerate.

In order to consider a dynamics on $\Sigma$, let us make the following
identifications:
\be
\left\{
\begin{array}{c}
y_0=q_2\\
y_3=p_2
\end{array}
\right.
\ \
\left\{
\begin{array}{c}
y_1=q_1\\
y_2=p_1
\end{array}
\right.
\ \ \ , \label{I1}
\ee
let us also define
\bea
H_i& =&p_i^2 + q_i^2~~~~~~~~~~~ i=1,2 \nonumber\\
H& =&H_1+H_2 \label{II1}
\eea
it follows that ${\cal C}_2= H$.
Moreover, using the fact that bivector fields defined on a 2-dimensional
surface can, at least locally, be written as $\Lambda=X \wedge Y$,
the Poisson bivector field given by equation \eqn{Lambdacoord} may be expressed
as:
\be
\Lambda=\left(\Delta-{H\over H_2} \Delta_2\right) \wedge\Gamma_2
\label{Lambdadin}
\ee
where $\Delta$ is the dilation vector field in $\R^4$ written as:
\bea
\Delta & = & \Delta_1+\Delta_2 \nonumber\\
\Delta_i & =& q_i\del_{q_i} + p_i\del_{p_i} \ \ \ i=1,2
\eea
and $\Gamma_2$ is defined as:\footnote{This vector field may be recognised
as the one of a 1-dimensional harmonic oscillator.}
\be
\Gamma_2=p_2\del_{q_2}-q_2\del_{p_2}.
\ee

Notice that decomposition \eqn{Lambdadin} is not unique, in fact $\Lambda$
may be written as $\Lambda=(a_1 X + b_1 Y) \wedge (a_2 X + b_2 Y)$, with
$$
L_X{\cal C}_1=L_X{\cal C}_2=0
$$
$$
L_Y{\cal C}_1=L_Y{\cal C}_2=0
$$
and
$$
X\wedge Y\neq 0, \ \ a_1 b_2 - a_2 b_1 = 1,
$$
so that there is a 3-parameter family of different decompositions of the
bivector field \eqn{Lambdadin}.

Identifications \eqn{I1}, \eqn{II1}, suggest to regard $H_1$ and $H_2$ as
oscillator Energies. We can use our bracket to associate with them equations
of motion, the resulting system is composed by two non interacting
oscillators, let us say $1$ and $2$, whose total energy is fixed by the
Casimir ${\cal C}_2$. Moreover the Casimir 1-form $\alpha _1$ may be
recognized to be the differential of ${\phi}_1$, the phase of oscillator
$1$, up to a multiplicative factor: $\alpha _1 = p_1 dq_1 - q_1 dp_1 =
H_1d\phi_1$. The equations of the motion are, for oscillator $2$
\bea
\dot q_2 = \{q_2,H_2\} & = & -2 H_1 p_2 \nonumber\\
\dot p_2 = \{p_2,H_2\} & = & 2 H_1 q_2 \nonumber\\
\{q_1,H_2\} =  \{p_1,H_2\} & = & 0 \label{O2}
\eea
and for oscillator $1$
\bea
\dot q_1 = \{q_1,H_1\} & = & 0 \nonumber\\
\dot p_1 = \{p_1,H_1\} & = & 0 \nonumber\\
\{q_2,H_1\} =  \{p_2,H_1\} & = & 0 \label{O1} \ \ .
\eea
As we may observe $p_1$ and $q_1$ are `frozen' degrees of freedom having
vanishing Poisson brackets with $H_1$ and $H_2$, so that oscillator $1$
does not evolve, but only furnishes the parameter $H_1$, which is in
fact a constant of the motion for oscillator 2. We therefore have a family of
1--dimensional oscillators whose motion is determined by an external parameter.
We also observe that,
since $H_1=H-H_2$ and $H$ is a constant, the system can also be regarded as
a non--linear oscillator, in the sense that it has an energy dependent
frequency, these kind of oscillators have being considered in the quantum
group context \cite{mmsz}.

We may conclude that, although the original phase space is  4-dimensional,
the motion happens to be on a 2--dimensional surface, that is
the dynamical system described has one degree of freedom, and its reduced
phase space is defined by ${\cal C}_1$ and ${\cal C}_2$. Indeed it turns out
 to be an oscillator (the one we indicated as oscillator $2$) which can be
regarded either as a non linear one, (in the specified acception) or as a
parameter dependent oscillator, the parameter being furnished by a reference
oscillator (in our notation oscillator $1$); we could say in the second case
that the phase of one oscillator `controls' the other.
Therefore we learn that Poisson manifolds provide an appropriate setting to
deal with parameter depending systems, each system leaving on a
symplectic leaf of the given Poisson manifold.

\subsection{$SU(1,1)$}

In analogy with what has been done for $SU(2)$ we now consider a \lp structure
on $SU(1,1)$ and look for a dynamical system defined on the group. We represent
the elements of the group as complex matrices  with unit determinant:
\be
g = y^0 \I + i \tilde \sigma_j y^j
\ee
where $\I$ is the identity matrix and
$$\tau_i=i\tilde\sigma_i$$ are the infinitesimal generators of the
group, the $\tilde\sigma_i$'s being defined as:
\be
\tilde\sigma_1=
\left(
\begin{array}{cc}
0 & -i\\
-i & 0
\end{array}
\right )
\ \ , \ \
\tilde\sigma_2=
\left(
\begin{array}{cc}
0 & 1\\
-1 & 0
\end{array}
\right )
\ \ , \ \
\tilde\sigma_3=
\left(
\begin{array}{cc}
1 & 0\\
0 & -1
\end{array}
\right )
\ \ ;
\ee
moreover $\det g = y_0^2 - y_1^2 - y_2^2 + y_3^2=1.$
The left and right invariant vector fields are respectively
\bea
Y_{1 \pm} & = & y_0 \del_1 + y_1 \del_0 \pm (y_2 \del_3 + y_3 \del_2)
\nonumber\\
Y_{2 \pm} & = & y_0 \del_2 + y_2 \del_0 \mp (y_3 \del_1 + y_1 \del_3)
\nonumber\\
Y_{3 \pm} & = & y_0 \del_3 - y_3 \del_0 \mp (y_1 \del_2 - y_2 \del_1)
\label{coo2}
\eea
where $X_i=Y_{i+},~~\tilde X_{i}=Y_{i-}$.
Also in this case we consider Poisson structures of the form
\be
\Lambda=X_i\wedge X_j - \tilde X_i\wedge \tilde X_j \label{lamsu11}
\ee
which are in fact \lp structures, as already noted.
{}From a topological point of view the group manifold is a connected
hyperbolic surface in $\R^4$, that is an hypercilinder $S^1\times \R^2$.
For this reason the various bivector-fields \eqn{lamsu11} that we may consider
on $SU(1,1)$ are not equivalent. Basically, we have two inequivalent
situations: one characterised by taking the two `boosts', which in our
notation are $X_1$ and $X_2$ ($\tilde X_1, \tilde X_2$, in the right invariant
realization), the other characterised by taking the rotation vector field,
namely $X_3$ and its right invariant partner, $\tilde X_3$, and one of the two
`boosts'.
\vskip 1cm

\noindent{\bf$a)$ The `boost-boost' case }

Let us first consider the case where the vector fields involved in the
definition of the Poisson structure on $SU(1,1)$ are the two `boosts'.
In that case the bivector field is:
\be
\Lambda =
{1\over 2} (X_1 \wedge X_2 - \tilde X_1 \wedge \tilde X_2) \label{Lambda1} \ \
{}.
\ee
Using the $\R^4$ representation of the vector fields given by \eqn{coo2}
we have:
\be
\Lambda =- (y_3 \del_0 - y_0 \del_3) \wedge \tilde\Delta- (\det g) \del_0
\wedge \del _3 \label{Lambdacoord2}
\ee
where we have indicated with $\tilde\Delta$ the vector field:
$$
\tilde\Delta=y_0\del_{y_0}+y_3\del_{y_3}-y_1\del_{y_1}-y_2\del_{y_2}\ \ .
$$
This Poisson structure is degenerate, having two Casimirs:
\be
{\cal C}_1 = \frac{y_1}{y_2}
\ee
\be
{\cal C}_2 = \det g .
\ee
If we do the same identification of variables as in the previous example we
obtain
\be
{\cal C}_2=H=H_2-H_1
\ee
and the other Casimir is unchanged. We may decompose the bivector field
$\Lambda$ the
same way as before, that is
\be
\Lambda = (\Delta - {H\over H_2} \Delta_2)\wedge \Gamma_2   \label{lambdadin2}
\ee
with $\Gamma_2$, $\Delta$, $\Delta_i$, having the same meaning as in the
previous example.
In fact the vector fields $X=\Delta - {H\over H_2} \Delta_2$ and $Y=\Gamma_2$
still satisfy $L_X {\cal C}_i=L_Y {\cal C}_i = 0$ with $i=1,2$.
Let us look now at the equations of the motion in order to understand which
kind of dynamical system we are describing.
We have:
\bea
\dot q_2 & = \{q_2,H_2\} =& 2 H_1 p_2 \nonumber\\
\dot p_2 & = \{p_2,H_2\} = & - 2 H_1 q_2 \nonumber\\
 \{q_1,H_2\}  & = \{p_1,H_2\} = & \{H_1,H_2\} = 0  \nonumber\\
\dot q_1 & = \{q_1,H_1\} =& 0 \nonumber\\
\dot p_1 & = \{p_1,H_1\} =&0 \ \ .
\eea
Again, the system we are describing appears to be a pair of non interacting
oscillators (being $\{H_1, H_2\}=0$), one of the two not evolving, so that the
reduced system turns out to be a parameter dependent 1-dimensional oscillator,
or a non linear one, depending on the possibility of writing $H_1$ as $H-H_2$.
That is, we have obtained the same result as in the $SU(2)$ case. This result
is due to the fact that the reduced phase space is, even in this case, a circle
in the plane identified by coordinates $p_2$ and $q_2$. In fact the bivector
\eqn{lambdadin2} has a local Casimir $\cal C_1$, whose Casimir 1-form
$\alpha_1 = y_2 dy_1-y_1 dy_2$ may be written as:
\be
\alpha _1 = p_1dq_1 - q_1dp_1 .
\ee
This one-form defines a non degenerate surface (provided the other Casimir is
imposed) in the plane $(p_2,q_2)$. In this plane the reduced phase space is
defined by
\be
p^2_2 + q^2_2 = H + H_1
\ee
which is a circle in the $(p_2,q_2)$ plane.

This result should not be a surprise, for we know that a given dynamical
system may admit more than one Hamiltonian description. We are finding here
that this statement may be true also for parameter depending systems.
\vskip 1cm
\noindent{\bf$b)$ The `rotation-boost' case}

As already noted, there is essentially only another inequivalent bivector
field we may consider on $SU(1,1)$ which we expect to give rise to a
different situation: namely we choose as starting vector fields a `boost'
and the rotation. For instance
\be
\Lambda =\frac{1}{2}( X_2\wedge X_3 - \tilde{X_2}\wedge\tilde{X_3})
  \label{lambda3}
\ee
(an equivalent possibility is
$\Lambda = {1\over 2} (X_1\wedge X_3 - \tilde{X_1}\wedge\tilde{X_3}) $ ).
We can still write $\Lambda$ in the form:
\be
\Lambda =  (\Delta - {H\over H_2}\Delta_2)\wedge\Gamma_2
\ee
but we change the identification of the physical variables. We
choose:
\bea
y_2 =& q_1 ~~~~~~~~~ y_1 &= q_2 \\ \nonumber
y_3 =& p_1 ~~~~~~~~~ y_0 &= p_2 \label{I2}
\eea
With this choice
\be
{\cal C}_2 = H = p^2_2 + p^2_1 - q^2_2 - q^2_1 = H_1 + H_2
\ee
where
\bea
H_1 &=& p^2_1 - q^2_1 \\
H_2 &=& p^2_2 - q^2_2 .
\eea
The previous identification is suggested by the Casimir 1-form $\alpha_1$ which
in this case is:
\be
\alpha _1 = y_3dy_2 - y_2dy_3 .
\ee
The expression of the Casimir suggests that $y_2$ and $y_3$ should be conjugate
variables if we want to continue to think of ${\cal C}_1$ as a quantity
which inherits only one of two particles constituting our dynamical system.
Obviously other identifications will describe different dynamics and there
is no preferred choice.

 We observe that the hamiltonians $H_1$, $H_2$, are hamiltonians of 'inverted'
oscillators, that is particles rolling down on inverted parabolic
slope. As before, one 'oscillator' is controlled by the other, so that the
trajectory of the reduced system in the phase space will be an hyperbola. This
can be understood by the fact that $H_1$ (the Hamiltonian of the control
system) is a constant of the motion for the system $2$, so, considering the
Casimir ${\cal C}_2$, that is the total hamiltonian, we are left with:
\be
p^2_2 - q^2_2 = H - H_1
\ee
the equation of an hyperbola.
However, as in the previous examples, the 'frequency' of the inverted
oscillator depends on the energy of the other one.

\section
{Lie-Poisson Structures and Dissipative Fields}

The Lie Poisson structures we have considered, being constant combinations
of left and right invariant bivector fields, belong to
the class of quadratic Poisson brackets of $\R^4$, when we describe the 3-
dimensional group manifold under consideration as embedded in $\R^4$.
They are of the form
\be
\{y_i,y_j\} = R^{rs}_{ij}y_r y_s \label{quad}
\ee
with $R^{rs}_{ij} \in \R$ and $y_i, i=0,..3$ coordinates in $\R^4$.
We want to consider now some particular aspects of quadratic Poisson brackets.
The Poisson bracket \eqn{quad} is left invariant by the dilation vector field
\be
\Delta = y_j{\partial\over \partial y_j} \label{diss}.
\ee
$\Delta$ can be thought of as a prototype of dissipative field, and
thus, as dynamical vector field, is compatible with the Poisson bracket. On
the other hand a quadratic Poisson Bracket, being zero along the zero of the
quadratic form $R^{rs}_{ij}y_r y_s$, cannot be inverse of a symplectic
structure. Therefore any quantization procedure relying on the existence of
a symplectic structure cannot be applied in the present situation. However,
since the deformed product is itself invariant under $\Delta$, at least for
the two Lie groups we have considered \cite{tak},
$*$-quantization procedure can still be used. So, in these cases,
dissipative dynamical systems survive the transition to quantum groups, or
they survive the quantization procedure.
In the coming examples we will consider some
dynamical systems preserving the Lie-Poisson group structure.

The form of the Poisson bivector fields considered until now, namely
$$
\Lambda =  (\Delta - {H\over H_2}\Delta_2)\wedge\Gamma_2
$$
suggests the investigation of dynamical fields of the form
\be
\Gamma= {H \over H_2} \Delta_2 - \Delta \ \ \label{fieldGamma} .
\ee
These vector fields, are compatible with the \lp
structures under consideration, that is
\be
L_\Gamma \Lambda =0 \ \
\ee
so we are in the case we discussed above.
As we have already mentioned, these dynamical fields may be automorphisms
of the deformed (by $*$-quantization) algebra of observables, depending on
whether or not the deformed product is a function of the Poisson bracket.
Let us explain better this point. The commutator defined via the
$*_\hbar$-product ($*_\hbar$ means that the $*$ may be written as a series
of powers in $\hbar$)
$$
[f, g] = f *_\hbar g - g *_\hbar f
$$
is equal to the Poisson bracket only in the limit $\hbar \rightarrow 0$. Then
automorphisms of the Poisson bracket will not be  in general automorphisms of
the commutator. Nonetheless if the $*$-product is a function of the Poisson
bracket as in the $\R^{2n}$ case (in that case Weyl quantization furnishes $f *
g = \exp (\frac {\hbar}{2} \Lambda) (df \wedge dg)$, then any automorphism of
the Poisson bracket also preserves the commutator. For our examples $SU(2)$
and $SU(1,1)$ we may use the duality between the algebra of functions, ${\cal
F}(G)$ on a Lie
group and the universal enveloping algebra $U(\g)$ of the Lie algebra.
Namely, given the product, $\cdot$, on ${\cal F}(G)$ and the coproduct $\Delta$
on $U(\g)$ we have
\be
<a\cdot b,\xi> = <a\otimes b,\Delta(\xi)>~~~~~~~ a,b\in{\cal F}(G),~~~~\xi\in
\g
\ee
where $<a\cdot b,\xi>$ has to be understood as the Lie derivative of $a \cdot
b$ with respect to $X_{\xi}$, the realization of $\xi$ on the group;
analogously the right hand side is  $X_{\xi}(a) X_{\xi}(b)$, where we have
used the expression of the coproduct for the generators, namely
$\Delta (\xi)=\xi \otimes 1+ 1 \otimes \xi$.
The same relation holds for the deformed algebras \cite{tin}:
\be
<a* b,\xi> = <a\otimes b,\Delta_\hbar(\xi)>\ \ .
\ee
This means that whenever the deformed coproduct $\Delta_\hbar$ is invariant,
so is the deformed product $*$.
For our examples the coproduct is:
\bea
\Delta_\hbar(X^3) & = & X^3 \otimes 1 + 1 \otimes X^3 \nonumber \\
\Delta_\hbar(X^\pm) & = & X^\pm \otimes e^{-{\hbar\over 2} X^3}
+  e^{{\hbar\over 2} X^3} \otimes X^{\pm}
\eea
where $X^\pm,X^3$ are the generators of the algebra. It can be verified that
this coproduct has the same invariance properties as the Poisson bracket
defined on the algebra of functions.

These considerations allow the
investigation in a quantum setting of dynamical fields which preserve \lp
structures of $SU(2)$ and $SU(1,1)$.
The examples we are going to describe are again only $SU(2)$ and $SU(1,1)$,
however we will not consider the quantization problem.

\subsection{ $SU(2)$}

We use in this example the identification of variables of example $(3.1)$ and
we consider the dynamical vector field \eqn{fieldGamma}. In order to
understand the dynamics we evaluate the equations of the motion which turn
out to be
\be
\begin{array}{rcccl}
\dot q_1 & = & i_\Gamma dq_1 & = & - q_1\\
\dot p_1 & = & i_\Gamma dp_1 & = & - p_1\\
\dot q_2 & = & i_\Gamma dq_2 & = &  {H_1 \over H_2} q_2\\
\dot p_2 & = & i_\Gamma dp_2 & = & {H_1 \over H_2} p_2
\end{array} \label{eqmotsu2}
\ee
Obviously the Casimirs ${\cal C}_1$, ${\cal C}_2$ are still  constants of
motion, so that
the motion is on the 3-sphere of constant radius $H_1+H_2$. It is
useful to visualize the motion in the two planes $(q_1,p_1)$ and $(q_2,p_2)$,
which can be considered as the phase spaces of two particles. The motion is
described in figures 1
\begin{figure}
\vspace{9.5cm}
\caption{\sl  The section of the flow diagram for the $SU(2)$ case
in the $p_1,q_1$ plane.}
\end{figure}
and 2.
\begin{figure}
\vspace{9.5cm}
\caption{\sl  The section of the flow diagram for the $SU(2)$ case in
the $p_2,q_2$ plane.}
\end{figure}
For both particles (because of the
constraint) the motion is confined in the disk of radius $H$, with the
center of one circle  corresponding to the boundary of the other.

For particle 1 the motion is a contraction to the center, independently of
the initial condition, with radial velocity proportional to the radius.
The center of the $(p_1q_1)$ circle is therefore a sink. By evolving back in
time
it is easy to see that the boundary of the disk is a source, that is the
trajectory of a particle can start from it at a finite time.
For the second particle the situation is
reversed, the motion is now an expansion, but with a speed which goes to 0
as $t\to\infty$, so now the center is a source, and boundary a sink. From
the global $S^3$ point of view we have that the two--sphere $p_1^2+q_1^2=1$
is the sink, while the two-- sphere $p_2^2+q_2^2=1$ is the source. Obviously
reversing the sign of the field will change the sink to a source and
viceversa.

\subsection{ $SU(1,1)$}

As we have seen, in the $SU(1,1)$ case there are two inequivalent \lp
structures which we expect to give rise to different
dynamics even in the case we are analyzing.
In both situations we consider the dynamical field to be \eqn{fieldGamma},
but using the two different identifications of variables \eqn{I1} and \eqn{I2}.
Let us first consider the \lp structure defined in \eqn{Lambda1} and the
identifications in \eqn{I1}. The equations of the motion are then

\bea
\dot q_1 & = &  q_1\nonumber\\
\dot p_1 & = & p_1\nonumber\\
\dot q_2 & = & {H_1 \over H_2} q_2\nonumber\\
\dot p_2 & = & {H_1 \over H_2} p_2 \label{eqmotsu111}
\eea

This time the motion is on a one--sheeted hyperboloid in $\R^4$. We indicate
it in figures 3
\begin{figure}
\vspace{9.5cm}
\caption{\sl  The section of the flow diagram for the $SU(1,1)$
`boost--boost' case in
the $p_1,q_1$ plane.}
\end{figure}
and 4
\begin{figure}
\vspace{9.5cm}
\caption{\sl  The section of the flow diagram for the $SU(1,1)$
`boost--boost' case in
the $p_2,q_2$ plane.}
\end{figure}
depicting the situation, note that while the
variables $p_1$ and $q_1$ can take any value, due to the constraint, $p_2$
and $q_2$ can only take values outside of the unit circle of the $(p_2,q_2)$
plane. The motion of the first particle is a radial expansion, with speed
proportional to the radius, while particle 2 has a source on the boundary
of its phase space, and then it expands with a speed which increases with
time.

For the next case we consider the \lp structure
$$
\Lambda = X_2 \wedge X_3 - \tilde X_2 \wedge \tilde X_3
$$
and identify the variables of the phase space as in \eqn{I2}.
This time $H_i=p_i^2 - q_i^2$, and $H=H_1+H_2$. With these choices the
dynamical vector field gives the following equations of the motion
\bea
\dot q_1 & = & - q_1\nonumber\\
\dot p_1 & = & - p_1\nonumber\\
\dot q_2 & = & {H_1 \over H_2} q_2\nonumber\\
\dot p_2 & = & {H_1 \over H_2} p_2 \label{eqmotsu112}
\eea

Although apparently similar to the equations in \eqn{eqmotsu2} and
\eqn{eqmotsu111}, the phase diagram is actually quite different as can be
seen from figure 5
\begin{figure}
\vspace{9.5cm}
\caption{\sl  The section of the flow diagram for the $SU(1,1)$
`boost--rotation' case in
the $p_1,q_1$ plane.}
\end{figure}
and 6.
\begin{figure}
\vspace{9.5cm}
\caption{\sl  The section of the flow diagram for the $SU(1,1)$
`boost--rotation' case in
the $p_2,q_2$ plane.}
\end{figure}
In the $(q_1p_1)$ plane the motion is a simple
contraction with speed proportional to the radius, but in the $(q_2p_2)$
plane is rather more complex. The equilater hyperbola $p_2^2-q_2^2=1$
corresponds to the origin of the $(q_1p_1)$ plane, and is therefore a sink,
there are 6 different regions, on some of which the motion is contraction ,
on some a dilation (as depicted by the arrows in figure 6), the asymptots
of the hyperbola are equilibrium positions (they correspond to $H_1=0$), and
the origin is an hyperbolic saddle point.

\section{Conclusions}
The procedure illustrated for the simple cases of $SU(2)$ and $SU(1,1)$ may
be generalized to other Lie groups. That is one chooses a \lp
structure on a Lie group, and investigates the possible Hamiltonian dynamics
furnished by the Poisson bracket. This way we are provided with a wide
class of non trivial dynamical systems whose phase space is a \lp group.
This observation enables us to quantize using the deformation procedure which
leads to Quantum Groups. In many cases a canonical quantization of the
system under consideration is already available, nonetheless there may be
systems where this scheme could be a useful one.
 On the other side there is no standard quantization procedure which allows the
investigation of dissipative systems in a quantum setting so that the
$*$-deformation tool and quantum groups frame could be useful in the analysis
of the quantum meaning of such non conservative dynamics.
\vskip 1.5cm
\noindent {\bf AKNOWLEDGEMENTS}

We thank professor W.Thirring and professor P. Michor for their kind
hospitality at the Erwin Schroedinger Institut in Wien, where this work
was partially done.

% A macro to raise things. Used in math and journal macros.
\def\up#1{\leavevmode \raise.16ex\hbox{#1}}
%journal references
\newcommand{\npb}[3]{{\sl Nucl. Phys. }{\bf B#1} \up(19#2\up) #3}
\newcommand{\plb}[3]{{\sl Phys. Lett. }{\bf #1B} \up(19#2\up) #3}
\newcommand{\revmp}[3]{{\sl Rev. Mod. Phys. }{\bf #1} \up(19#2\up) #3}
\newcommand{\sovj}[3]{{\sl Sov. J. Nucl. Phys. }{\bf #1} \up(19#2\up) #3}
\newcommand{\jetp}[3]{{\sl Sov. Phys. JETP }{\bf #1} \up(19#2\up) #3}
\newcommand{\rmp}[3]{{\sl Rev. Mod. Phys. }{\bf #1} \up(19#2\up) #3}
\newcommand{\prd}[3]{{\sl Phys. Rev. }{\bf D#1} \up(19#2\up) #3}
\newcommand{\ijmpa}[3]{{\sl Int. J. Mod. Phys. }{\bf A#1} \up(19#2\up) #3}
\newcommand{\prl}[3]{{\sl Phys. Rev. Lett. }{\bf #1} \up(19#2\up) #3}
\newcommand{\physrep}[3]{{\sl Phys. Rep. }{\bf #1} \up(19#2\up) #3}
\newcommand{\journal}[4]{{\sl #1 }{\bf #2} \up(19#3\up) #4}

%
%%%%%%%%%%%%%%%%%%%%%%%%%%%%%%%%%%%%%%%%%%%%%%%%%%%%%%%%%%%%%%%
%
%      List of references
%
%%%%%%%%%%%%%%%%%%%%%%%%%%%%%%%%%%%%%%%%%%%%%%%%%%%%%%%%%%%%%%%%
%

\end{document}